%% file: main.tex
\documentclass[conference]{IEEEtran}
\IEEEoverridecommandlockouts
\usepackage{amsmath,amssymb,amsfonts}
\usepackage{algorithmic}
\usepackage{graphicx}
\usepackage{textcomp}
\usepackage{xcolor}

\newcommand{\Real}{\mathbb{R}}
\newcommand{\BV}{\mathbb{V}}

\newcommand{\BC}{\mathbb{C}\mathrm{ov}}
\newcommand{\BE}{\mathbb{E}}
\newcommand{\Prob}{\mathrm{Pr}}

\newcommand{\eps}{\epsilon}
\newcommand{\ordo}{\mathcal{O}}

\usepackage[backend = biber, sorting=none, style=ieee]{biblatex}
\usepackage{subfiles}
\usepackage{tikz}
\usetikzlibrary{arrows,patterns}
\def\BibTeX{{\rm B\kern-.05em{\sc i\kern-.025em b}\kern-.08em
    T\kern-.1667em\lower.7ex\hbox{E}\kern-.125emX}}

\begin{document}

\title{Analysis of Quantization Noise Suppression Gains in Digital Phased Arrays
\thanks{The research was funded by the strategic innovation program Smarter Electronics System, Vinnova, Sweden.}
}

\author{\IEEEauthorblockN{Erik Kennerland}
\IEEEauthorblockA{\textit{BeammWave AB} \\
223 70 Lund, Sweden \\
erik.kennerland@gmail.com}
\and
\IEEEauthorblockN{Bengt Lindoff}
\IEEEauthorblockA{\textit{BeammWave AB} \\
223 70 Lund, Sweden \\
bengt@beammwave.com}
\and
\IEEEauthorblockN{Emil Björnson}
\IEEEauthorblockA{\textit{KTH Royal Institute of Technology} \\
100 44 Stockholm, Sweden \\
emilbjo@kth.se}
\and
}

\maketitle

\begin{abstract}
Digital phased arrays have often been disregarded for millimeter-wave communications since the analog-to-digital converters (ADCs) are power-hungry. In this paper, we provide a different perspective on this matter by demonstrating analytically and numerically how the ADC resolution can be reduced when using digital phased arrays.
We perform a theoretical analysis of the quantization noise characteristics for an OFDM signal received and processed by a digital phased array, using Gaussian approximation of the OFDM signal. In particular, we quantify the quantization noise suppression factor analytically and numerically. This factor describes how much the coherent combining reduces the quantization noise as a function of the number of antennas, which allows for reducing the ADC bit resolution. For instance in a 8-16 antenna digital phased array the ADC resolution can be reduced with 1-2 bits compared to the ADC required for an analog phased array.
\end{abstract}


\section{Background}
\label{background}

The tremendous success of cellular communications continuously leads to capacity problems, which require new deployments and radio technologies. In 5G, the millimeter-wave (mmW) frequency ranges have been added to greatly expand the available spectrum and, thereby, the capacity. However, due to the poor radio propagation properties at mmW frequencies, the beamforming gains provided by phased arrays are also needed on the mobile device side. Currently, only analog phased array transceivers are on the market, while it is well-known that digital phased arrays have better communication performance. The digital architecture has been envisaged to be too complex for mass-market mobile devices. For instance, digital phase arrays need one analog-to-digital converter (ADC) pair (I and Q branch) per antenna in contrast to only one ADC pair for an analog phased array. The reason for such claims is that the ADC has been seen as a power-hungry component, especially for the high numbers of bits that are needed in an analog phased array architecture to limit the quantization noise and enable high-capacity communications.

In this paper, we analyze the quantization noise characteristics for a Gaussian signal received and processed by a digital phased array. We show that quantization noise suppression is achieved due to the coherent combining of the signals received by the antenna array. We furthermore derive an interpretable approximation of the quantization noise processing gain as a function of the number of antennas in the array and ADC bit resolution. The results show that the ADC design for digital phase arrays can be relaxed compared to analog counterparts, when it comes to performance impact due to quantization noise, thereby enabling possibilities for also using digital phase array architectures in low-cost mmW devices. 

\subsection{Related work}

ADC quantization for one-dimensional Gaussian signals was analyzed in \cite{VarianceQuant}. It was noted that a reduction in ADC resolution by 1 bit adds approximately 5 dB of quantization noise, which is slightly less than the classical rule of thumb of 6 dB that holds for the quantization of a sinus wave. 
The achievable data rates with low-bit ADCs have been studied in recent years \cite{Zhang2018a,LowResMIMO}, with a particular focus on mmW systems. It was demonstrated in \cite{Studer2016a,Mollen2017a} that a handful of bits per ADC is sufficient to reach close to the infinite-resolution case when there are many antennas. The intuition is that quantization noise averages out among the many antennas, but this phenomenon is generally overestimated by assuming independent quantization noise. Since the ADCs quantize correlated signals, the quantization noise is also correlated \cite{BjornsonHWdistortionCorrelation}. 
The need for studying the impact of quantization noise correlation was recognized in \cite{LowResMIMO}, and this research gap has not been filled yet. The goal of this paper is to do that.

\subsection{Organization}

In Section \ref{quantizationAnalysisSection}, we analyze the correlation properties of the ADC quantization noise of bivariate Gaussian signals and derive the convergence rates as the number of ADC bits $k\rightarrow \infty$. In Section \ref{phasedArrayAnalysisSection}, we utilize these results for determining analytical quantization correlation expressions for the ADC quantization noise of an OFDM signal received through a digital phased array. .In Section \ref{bounds}, we determine interpretable approximations for the quantization noise suppression as a function of the number of ADC bits and number of antennas. Section \ref{results} compares simulation with the determined analytical expression and approximations, while in Section \ref{implications}, we describe the implication of these results on the design of a mmW digital phase array architecture in mobile devices. The conclusions are stated in Section \ref{conclusions}.

\section{ADC Quantization Error Properties of Bivariate Gaussian Signals}
\label{quantizationAnalysisSection}

A focal objective of this paper is to determine the covariance between the quantization errors of complex Gaussian variables that differ by a phase shift, which happens when using phased arrays when receiving OFDM signals. Here we will derive analytical results that will then be used in Section~\ref{phasedArrayAnalysisSection} to achieve our objective.

We consider a time-continuous real-valued signal  $ x $ that is sampled by an ADC. By defining an equidistant grid $x_m = -R+(m-\frac{1}{2})\Delta$ for $1 \leq m \leq 2^k$ over $[-R,R]$,
a $k$-bit quantizer $Q$ with midrise characteristic and clipping level $R$ is defined as
\begin{equation}
Q(x) = \begin{cases}
     -x_s, & x \leq -R+\Delta,\\
     x_m, & x_m-\frac{\Delta}{2} \leq x < x_m+\frac{\Delta}{2},\\
    x_s, & R-\Delta < x,\\
\end{cases}
\end{equation}
where $\Delta = \frac{2R}{2^k}$ is the step size between the quantization levels, and $x_s \triangleq x_{2^k} = -x_{1} = R-\frac{\Delta}{2}$ is the saturation level. The quantization error $x-Q(x)$ is called the quantization noise.

\subsection{Optimal Clipping Level}
If a one-dimensional Gaussian signal with mean zero and unit variance is quantized, its quantization noise variance can be tightly approximated by 
\begin{equation} \label{eq:Vq}
  V_q \approx \frac{1}{3\cdot 2^{2k}}R^2+\sqrt{\frac{8}{\pi}}R^{-3}e^{-\frac{R^2}{2}}
\end{equation}
for clipping levels $ R $ around the clipping level maximizing the Signal to Quantization noise level (SQNR) and above, see \cite{VarianceQuant}.
It is straightforward to use this formula for other mean and variances by normalizing the signal before quantization.
    The optimal clipping level $R$ that minimizes the quantization error variance, for a fixed number of bits $k$, is therefore found by differentiating \eqref{eq:Vq} and equating it to zero, which after some elementary algebra simplifies to the equation
    \begin{equation}
    \label{optClipping}
    e^{-\frac{R^2}{2}} = \sqrt{\frac{\pi}{8}}\frac{2}{3}2^{-2k}\frac{R^5}{3+R^3} \in \ordo(R^2\,2^{-2k}).\end{equation}
Although there is no analytic expression, the equation can be easily solved numerically.
     Expressing the proportionality between $e^{-\frac{R^2}{2}}$ and $k$ as in \eqref{optClipping} is useful because of its relation to the often occurring complementary cumulative (Gaussian) distribution function $\Phi^C(x) \triangleq \Prob[X \geq x]$ evaluated at $R$. Particularly, note that for $x > 0$, \begin{multline}
    \label{CCDFapprox}
    \Phi^C(x) = \int_{x}^{\infty} \frac{1}{\sqrt{2\pi}}e^{-\frac{s^2}{2}}\,ds \leq \frac{1}{\sqrt{2\pi}}\int_{x}^\infty \frac{s}{x}e^{-\frac{s^2}{2}}\,ds = \\
    \frac{\phi(x)}{x} \in \ordo(x^{-1}e^{-\frac{x^2}{2}}),
    \end{multline}
    where $\phi$ denotes the univariate unit Gaussian probability density function.
    The correspondence $e^{-\frac{R^2}{2}} \in \ordo(R^22^{-2k})$ will be used in later sections to determine the convergence rates of quantization errors. At this point, we notice that (\ref{optClipping}) implies that $R \in \ordo(\sqrt{k})$ for large $k$. In the remainder of this paper, the clipping level is always assumed to be optimally chosen for a given $k$.

\subsection{Quantization Noise Covariance}

\subfile{Graphs/partitionGray.tex}

When comparing the quantization noise across antennas in a phased array, finding an applicable approximation to the covariance between quantized real-valued Gaussian distributed variables is of particular utility. We will consider the jointly Gaussian distributed variables $(X,Y)$ with the joint probability density function $\varphi(x,y;\rho)$, where $\rho$ is the Pearson-correlation coefficient between the marginal (Gaussian) distributions $X$ and $Y$. Their quantization errors $\eps_X^q = X-Q(X)$ and $\eps_Y^q = Y-Q(Y)$ yield the real-valued quantization error covariance $\psi$:
\begin{align}
\nonumber
    \psi(\rho)&\triangleq\BC[\eps_X^q,\eps_Y^q] \\ &=\int_{\Real^2}(x-Q(x))(y-Q(y))\varphi(x,y;\rho)\,dxdy. \label{psiDefinition}
\end{align}
To evaluate this integral, we partition $\Real^2$ into the rectangular regions where $Q(x)$ and $Q(y)$ are constant, as shown in Fig.~ \ref{partition}. We let $\mathcal{I}_{m,n} = [x_m-\frac{\Delta}{2},x_m+\frac{\Delta}{2}) \times [y_n-\frac{\Delta}{2},y_n+\frac{\Delta}{2})$ denote the interior regions for $1 \leq m,n \leq 2^k$. Secondly, let $\mathcal{C}_1 = [R,\infty) \times [R,\infty) \cup (-\infty, -R] \times (-\infty, -R]$ and $\mathcal{C}_2 = (-\infty,-R] \times [R,\infty) \cup [R, \infty) \times (-\infty, -R]$ denote the corner regions for which both $X$ and $Y$ saturate. Lastly, let $\mathcal{E} = \mathbb{R}^2\setminus (\mathcal{C}_1 \cup \mathcal{C}_2\cup \bigcup_{m,n = 1}^{2^k}\mathcal{I}_{m,n})$ be the edge region for which one, but not both, of $X$ or $Y$ are saturated. Separating the integration over the partition yields
\begin{align} \nonumber
\psi(\rho) = &\sum_{m,n= 1}^{2^k} \int_{\mathcal{I}_{m,n}}(x-x_m)(y-y_n)\varphi(x,y;\rho)\,dxdy 
\\ \nonumber &+\int_{\mathcal{C}_1}(x-x_s)(y-y_s)\varphi(x,y;\rho)\,dxdy
\end{align}
\begin{align}
 &+\int_{\mathcal{C}_2}(x+x_s)(y+y_s)\varphi(x,y;\rho)\,dxdy\nonumber\\
&+\int_{\mathcal{E}}(x-Q(x))(y-Q(y))\varphi(x,y;\rho)\,dxdy.\label{eq:psirho1}
\end{align}
The integration over $\mathcal{E}$ is of order $\ordo(R\,2^{-k}e^{-\frac{R^2}{2}})$ for all $\rho$. The precise proof of this claim is omitted for conciseness, but the main property is that the integration over the region corresponding to either $X$ or $Y$ decreases in proportion to the quantization step size $\Delta = \frac{2R}{2^k} \in \ordo(R2^{-k})$ because of the nonsaturating values, while simultaneously decreasing at the rate of $\phi(R) \in \ordo(e^{-\frac{R^2}{2}})$ as a consequence of the appearance of a one-sided truncation of a Gaussian distribution at $R$. This is in contrast to the integration over $\mathcal{C}_1$ and $\mathcal{C}_2$, which will later be proven to converge to zero at a rate of $\ordo(R^{-1}e^{-\frac{R^2}{2}})$ as $k \to \infty$. Using the notation $I_{m,n} = \int_{\mathcal{I}_{m,n}} (x-x_m)(y-y_n)\varphi(x,y;\rho)\,dxdy$, \eqref{eq:psirho1} simplifies into
\begin{multline*}
\psi(\rho) = \\2\int_{R,R}^{\infty,\infty} (x-x_s)(y-y_s)\big\{\varphi(x,y;\rho)- \varphi(x,y;-\rho)\big\}\,dxdy\\
+ \sum_{m,n = 1}^{2^k}I_{m,n} + \ordo(2^{-k}e^{-\frac{R^2}{2}}).
\end{multline*}
For $|\rho|$ sufficiently far from $1$, $|\rho| <1- 2^{-2k}$ for example\footnote{Strictly speaking, one needs to choose $1 - 2^{-tk}$ for some $0 < t < 2$. However, letting $t = 2$, instead of choosing an arbitrary $t$ close to $2$, makes the coming sections more legible.}, the contribution from the interior region vanishes much quicker than that of the corner region $|\sum_{m,n = 1}^{2^k} I_{m,n}| \ll \ordo(R^{-1}e^{-\frac{R^2}{2}})$. The detailed proof of this is also rather technical, but follows by considering the Taylor expansion of $\varphi$ around $(x_m,y_n)$ of each $\mathcal{I}_{m,n}$ and integrating. As $|\rho| \to 1$, however, the covariance collapses into the univariate quantization noise variance\footnote{This is identified by viewing the limit of $\varphi$ as $|\rho| \to 1$ in a distributional sense.} $\sigma^2$. Then, by denoting $C(\rho) = 2\int_{R,R}^{\infty,\infty} (x-x_s)(y-y_s)\varphi(x,y;\rho)\,dxdy$ the covariance is, up to $\ordo(R\,2^{-k}e^{-\frac{R^2}{2}})$ aptly approximated as
\begin{equation}
\label{psiApprox}
    \psi(\rho) = \begin{cases}
    C(\rho) - C(-\rho), & |\rho| -1 \leq 2^{-2k},\\
    \sigma^2,& |\rho| -1 > 2^{-2k}.\\
    \end{cases}
\end{equation}
By expanding $C(\rho)$, the expression is identified as the second- and first-order moments of truncated bivariate Gaussian distributions, which when utilizing their recursive description from \cite{truncatedBothSides}, results in
\begin{multline}
\label{cornerFunction}
    C(\rho) = \left((R-\frac{\Delta}{2})^2+\rho\right)\Prob[(X,Y) \in \mathcal{C}_1] + 2(1-\rho^2)\cdot\\ \varphi(R,R;\rho) -
    4\left(R-(1+\rho)\frac{\Delta}{2}\right)\phi(R)\Phi^C\left(\frac{R(1-\rho)}{\sqrt{1-\rho^2}}\right).
\end{multline}
Barring $\Phi^C$ and $\Prob[(X,Y) \in \mathcal{C}_1]$, we have thus established a closed-form approximation of $\psi$ which holds up to $\ordo(R\,2^{-k}e^{-\frac{R^2}{2}})$ for $|\rho|$ reasonably far from $1$. The numerical evaluation of $\Phi^C$ is a well-explored topic. For instance, the approximation (\ref{CCDFapprox}) may be used as it gains accuracy for greater positive values. Most approximations to $\Prob[(X,Y) \in \mathcal{C}_1]$ (see e.g. \cite{ALBERS199487}) rely on Taylor expansion methods whose accuracy depends on $\rho$. As such, $\Prob[(X,Y) \in \mathcal{C}_1]$ can be calculated through numerical integration of the bivariate probability density $\varphi(x,y;\rho)$ over $\mathcal{C}_1$.

\section{Quantization Error Covariance Analysis for Digital Phased Arrays}
\label{phasedArrayAnalysisSection}

In this section, we will study the covariance between complex Gaussian signals when a digital phased array receives a signal from a specific direction. 

\subsection{Two-Element Phased Array}
Consider a digital phased array consisting with two antennas. Let the complex-valued baseband representation of the received signals for the respective antennas be $s_1 = I+jQ$ and $s_2 = e^{j\alpha}s_1 = (I\cos\alpha -Q\sin\alpha) +  j(I\sin\alpha + Q\cos\alpha)$, where the phase shift $\alpha$ is dependent on the Angle of Arrival (AoA) $\theta$, the wavelength $\lambda$, and the distance $ d$ between the antennas.\footnote{It is $\alpha = -2\pi d \sin(\theta) / \lambda$ if the AoA is measured from the boresight.} The signals $ s_1, s_2$ are assumed to be standard circular-symmetric complex Gaussian random variables, which is a good approximation of OFDM signals \cite{OFDMGauss}. Let the respective I- and Q-branch of the signals $s_1, s_2$ go through a $k$-bit ADC. We denote the (complex) quantization noise of $s_1$ by $\epsilon_{C,1}^{q,k}$ and its variance by $2\sigma^2$. Because of the phase displacement, the quantization noise $\eps_{C,2}^{q,k}$ of $s_2$ will differ from that of $s_1$. We separate the I- and Q-branches of the quantization errors as $\eps_{C,1}^{q,k} = \eps_{I,1}^{q,k} + j\eps_{Q,1}^{q,k}$ and $\eps_{C,2}^{q,k} = \eps_{I,2}^{q,k} + j\eps_{Q,2}^{q,k}$. The separation into real-valued quantization noises allows for the application of $\psi$ in (\ref{psiApprox}), since $\psi(\rho)$ by definition is the covariance between the quantization noises of one-dimensional Gaussian signals between signals with known correlation $\rho$ before quantization. In this case, the correlation coefficients between the various I- and Q-branches are either $\cos \alpha$ or $\sin \alpha$:
\begin{align*}
    \BC[\eps_{I,1}^{q,k},\eps_{I,2}^{q,k}] &= \psi(\BC[I, I\cos\alpha - Q\sin\alpha])  = \psi(\cos\alpha), \\
    \BC[\eps_{I,1}^{q,k}, \eps_{Q,2}^{q,k}] &= \psi(\BC[I,I\sin\alpha+Q\cos\alpha]) = \psi(\sin\alpha), \\
\BC[\eps_{Q,1}^{q,k},\eps_{I,2}^{q,k}] &= \psi(\BC[Q,I\cos\alpha - Q\sin\alpha]) = -\psi(\sin\alpha), \\\BC[\eps_{Q,1}^{q,k},\eps_{Q,2}^{q,k}] &= \psi(\BC[Q,I\sin\alpha + Q\cos \alpha]) = \psi(\cos\alpha).
\end{align*}
To perform coherent combining of the received sampled signals $ s_1, s_2$, a reverted phase is applied to the second sampled signal, giving a phase-rotated quantization noise, $e^{-j\alpha}\eps_{C,2}^{q,k} = ( \eps_{I,2}^{q,k}\cos\alpha +  \eps_{Q,2}^{q,k}\sin \alpha) + j(\eps_{Q,2}^{q,k}\cos\alpha -  \eps_{I,2}^{q,k}\sin\alpha)$.
Omitting the superscripts $q$ and $k$ for brevity, we can expand the covariance between the received quantization errors as
\begin{multline}
r_\eps(\alpha) \triangleq\BC[\eps_{C,1}^{q,k},e^{-j\alpha}\eps_{C,2}^{q,k}] =\\
\BE[\eps_{I,1}\eps_{I,2}]\cos\alpha + \BE[\eps_{I,1}\eps_{Q,2}]\sin\alpha + \BE[\eps_{Q,1}\eps_{Q,2}]\cos\alpha -\\ \BE[\eps_{Q,1},\eps_{I,2}]\sin\alpha +
 j\big(\BE[\eps_{Q,1}\eps_{I,2}]\cos\alpha + \BE[\eps_{Q,1}\eps_{Q,2}]\sin\alpha -\\ \BE[\eps_{I,1}\eps_{Q,2}]\cos\alpha+ \BE[\eps_{I,1}\eps_{I,2}]\sin\alpha\big)\\
 = 2\cos\alpha\psi(\cos\alpha)+2\sin\alpha\psi(\sin\alpha) + \\2j\Big(\sin\alpha\psi(\cos\alpha) - \cos\alpha\psi(\sin\alpha)\Big).
\end{multline}
We have thus established the quantization noise covariance for a two-element phased arrays. Generally, there are more than $2$ elements in a phased array. However, as we will soon prove, this $2$-element expression may be used to construct the general quantization error covariance in phased arrays with $N$ antennas.

\subsection{Multiple Antenna Digital Phased Arrays}
We will now extend the analysis to consider an $N$-antenna uniform linear array (ULA). We define the corresponding complex-valued baseband signals $s_1,s_2,\ldots s_N$, where $s_1= I + jQ$ and $s_n = e^{j\alpha_n}s_1$ for $n \geq 1$. The quantization error covariance between an arbitrary pair $s_{m}$ and $s_{n}$ of signals depends only on the phase displacement difference, since $s_m = e^{j(\alpha_m - \alpha_n)}s_n$. For this reason 
\begin{align} \nonumber
   \BC\Big[e^{-j\alpha_n}\eps_{C,n}^{q,k}, e^{-j\alpha_m}\eps_{C,m}^{q,k}\Big] &=\BC\Big[\eps_{C,n}^{q,k}, e^{-j(\alpha_m-\alpha_n)}\eps_{C,m}^{q,k}\Big] \\
   & = r_\eps(\alpha_m-\alpha_n). 
\end{align}
After coherent combining, the total quantization error is the sum
\begin{equation}
\eps^q = \sum_{n = 1}^{N} e^{-j\alpha_n}\eps_{C,n}^{q,k},
\end{equation}
where $\eps_{C,n}^{q,k}$ is the quantization error of $s_n$. Its variance is
\begin{align} \nonumber
\BV[\eps^q] &= \sum_{m,n = 1}^N\BC[ e^{-j\alpha_n}\eps_{C,n}^{q,k}, e^{-j\alpha_m}\eps_{C,m}^{q,k}]
\\ &= \sum_{m,n = 1}^Nr_\eps(\alpha_m-\alpha_n).
\end{align}
For the antenna spacing $d$ in the ULA, the phase shifts becomes
$\alpha_n = -\frac{2\pi d}{\lambda}(n-1)\sin(\theta)$ when the signal arrives from the AoA
 $\theta$ and the wavelength is $\lambda$. In particular, they are translation invariant in the sense that $\alpha_{m}-\alpha_{n} = \alpha_{m-n+1}$, which may be used to simplify the total quantization variance as
\begin{multline}
\label{varianceExpression}
\BV[\eps^q] = \sum_{m,n = 1}^N r_\eps(\alpha_{m-n+1}) =
\sum_{n = -N+1}^{N-1} (N-|n|) r_\eps(\alpha_{n+1}) =\\ N r_\eps(\alpha_1) + \sum_{n = 1}^{N-1} (N-|n|)\big\{r_\eps(\alpha_{n+1})+r_\eps(-\alpha_{n+1})\big\} = \\
2\sigma^2 N + 4\sum_{n = 1}^{N-1} (N - n)\Psi(\alpha_{n+1}),
\end{multline}
where 
\begin{equation}
\label{boldPsi}
\Psi(\alpha) \triangleq \frac{1}{4}(r_\eps(\alpha) + r_\eps(-\alpha)) = \cos\alpha\psi(\cos\alpha) + \sin\alpha\psi(\sin\alpha).
\end{equation}
We note that if the quantization noises are uncorrelated for each pair of antennas, which can be seen as the ideal case, the coherently summed quantization noise yields a variance of $2\sigma^2 N$, corresponding to the first term in (\ref{varianceExpression}). Conversely, in case the quantization noise correlation between all antennas is $1$, (i.e., when $\Psi(\alpha_{n+1}) = \sigma^2$ for all $n$), the worst-case covariance is achieved at $2\sigma^2N^2$. This corresponds to the case when the quantization noise is also coherently combined, just as the received desired signals $ s_m$.

We now define the noise suppression factor as 
\begin{equation}
\label{gammaDef}
    \Gamma(\theta)=\frac{\BV[\eps^q(\theta)]}{2\sigma^2 N^2},
\end{equation} 
which measures how much smaller the total quantization noise variance is compared to the worst-case situation. The value ranges from $1$ to $\frac{1}{N}$, where a smaller value implies that the quantization noise originating from the ADC is suppressed due to the combining. As can be seen, the value depends on the AoA $ \theta $. To quantify the average noise suppression factor, we assume that the AoA $\theta$ is uniformly distributed in $(-\frac{\pi}{2},\frac{\pi}{2})$. This is a reasonable assumption because a digital phased array in a handheld mobile device can be rotated arbitrarily by the user, who is generally unaware of the base station location.
 Then, the average quantization noise suppression factor becomes
\begin{align}
\nonumber
&\BE_\theta (\Gamma(\theta)) \\ \nonumber
&=\frac{1}{\pi}\Bigg(\frac{\pi}{N} + \frac{2}{\sigma^2}\int_{-\frac{\pi}{2}}^{\frac{\pi}{2}} \sum_{n = 1}^{N-1}\frac{(N-n)}{N^2}\Psi \left(\frac{2\pi d}{\lambda}n\sin\theta \right) \,d\theta\Bigg) \\
&=\frac{1}{N} + \frac{2}{\pi\sigma^2} \sum_{n = 1}^{N-1}\frac{(N-n)}{N^2}\int_{-\frac{\pi}{2}}^{\frac{\pi}{2}}\Psi \left(\frac{2\pi d}{\lambda}n \sin\theta \right)\,d\theta. \label{suppression}
\end{align}

\subsection{Approximations of the Noise Suppression Factor}
\label{bounds}
The quantization noise suppression factor in \eqref{suppression} can be computed numerically, but to obtain more insights, we will also present two approximations of it.

First, the approximation in (\ref{psiApprox}) may be inserted directly into (\ref{boldPsi}) and then used to simplify the suppression function in (\ref{suppression}). This is useful for simulation purposes, but is unsuitable for hand-calculations. Therefore, we will develop a simpler approximation that, while perhaps less accurate, holds as a ``rule of thumb''. Since $C(\rho)$ is strictly increasing over $(-1,1)$, it is bounded above by $M \triangleq\lim_{\rho \to 1} C(\rho)$. Note further that $\lim_{\rho \to 1} C(-\rho) = 0$. Utilising $\Prob[(X,Y) \in \mathcal{C}_1] = \Prob[X \geq R] - \Prob[X \geq  R, Y \leq R] = \Phi^C(R) - \Prob[X\geq R, Y \leq R]$ the limit of the corner probability is $\Phi^C(R)$ as $\rho \to 1$, by using in \cite[Prop.~1]{ALBERS199487}. The remaining terms in (\ref{cornerFunction}) have trivial limits, giving
\begin{equation}
M = 2\left((R-\frac{\Delta}{2})^2+1\right)\Phi^C(R)-2(R-\Delta)\phi(R).
\end{equation}
This proves the claim from Section~\ref{quantizationAnalysisSection} that $C \in \ordo(R^{-1}e^{-\frac{R^2}{2}})$ for all $\rho$. A crude upper bound on $\psi(\rho)$ is thus 
\begin{equation}
\label{psiBound}
    |\psi(\rho)| \leq \begin{cases}
    M, & |\rho| \leq  1-2^{-2k},\\
    \sigma^2,& |\rho| > 1 - 2^{-2k}.\\
    \end{cases}
\end{equation}
Identifying local maximum points of (\ref{boldPsi}), it is not difficult to see that $\Psi \leq |\psi|$. In particular, if (\ref{psiBound}) is applied to $\Psi(\alpha_n) = \Psi(\frac{2\pi d}{\lambda}n\sin\theta)$, the upper value $\sigma^2$ is attained precisely at those $\theta$-values for which $|\cos \alpha_n| > 1-2^{-2k}$ or $|\sin \alpha_n| > 1-2^{-2k}$. Denote the subset of  $\theta$ for which this is the case by $U$ and $U^C = [-\frac{\pi}{2},\frac{\pi}{2}]\setminus U$ its complement. The upper bound of $\Psi$ may then simply be stated as $\Psi \leq \sigma^2 \chi_U+ M\chi_{U^C}$, where $\chi$ denotes a characteristic function. In other words, $$
\int_{-\frac{\pi}{2}}^{\frac{\pi}{2}}\Psi \left(\frac{2\pi d}{\lambda}n \sin\theta \right)\,d\theta \leq \sigma^2\int_{-\frac{\pi}{2}}^{\frac{\pi}{2}}\chi_U\,d\theta+ M \int_{-\frac{\pi}{2}}^{\frac{\pi}{2}} \chi_{U^C}\,d\theta.
$$
The measure of $U$, i.e., the probability that $|\cos \alpha_n|$ or $|\sin \alpha_n|$ is greater than $1-2^{-2k}$, is approximately $4\frac{\sqrt{2}}{\pi}\cdot 2^{-k} \approx 2\cdot 2^{-k}$. Hence, $\Psi \leq 4\frac{\sqrt{2}}{\pi}\cdot2^{-k}\sigma^2+(1-4\frac{\sqrt{2}}{\pi}\cdot2^{-k})M$, and this upper bound may be inserted  into (\ref{suppression}) to yield an convenient rule-of-thumb of the noise suppression, namely
\begin{multline}
    \label{upperBoundSuppression}
\BE_\theta[\Gamma] \leq
\frac{1}{N} + \frac{N-1}{N\sigma^2}\Bigg(4\frac{\sqrt{2}}{\pi}2^{-k}\sigma^2 + \Big(1-4\frac{\sqrt{2}}{\pi} 2^{-k}\Big)M\Bigg).
\end{multline}
As always, the approximate upper bounds holds only up to the order of the terms that are ignored in \eqref{psiApprox}. The first term in \eqref{psiApprox} corresponds to the noise suppression in case of uncorrelated quantization noise, while the latter term in does not decrease as $N$ increases, due to the fact that an increased number of antennas also leads to the risk of their phase shifts coinciding for various angles of arrival and hence gives a bound of the achievable quantization noise suppression, for a given $k$-bit ADC.

\section{Simulation Results}
\label{results}

\begin{figure}
    \centering
    \includegraphics[width=.85\columnwidth]{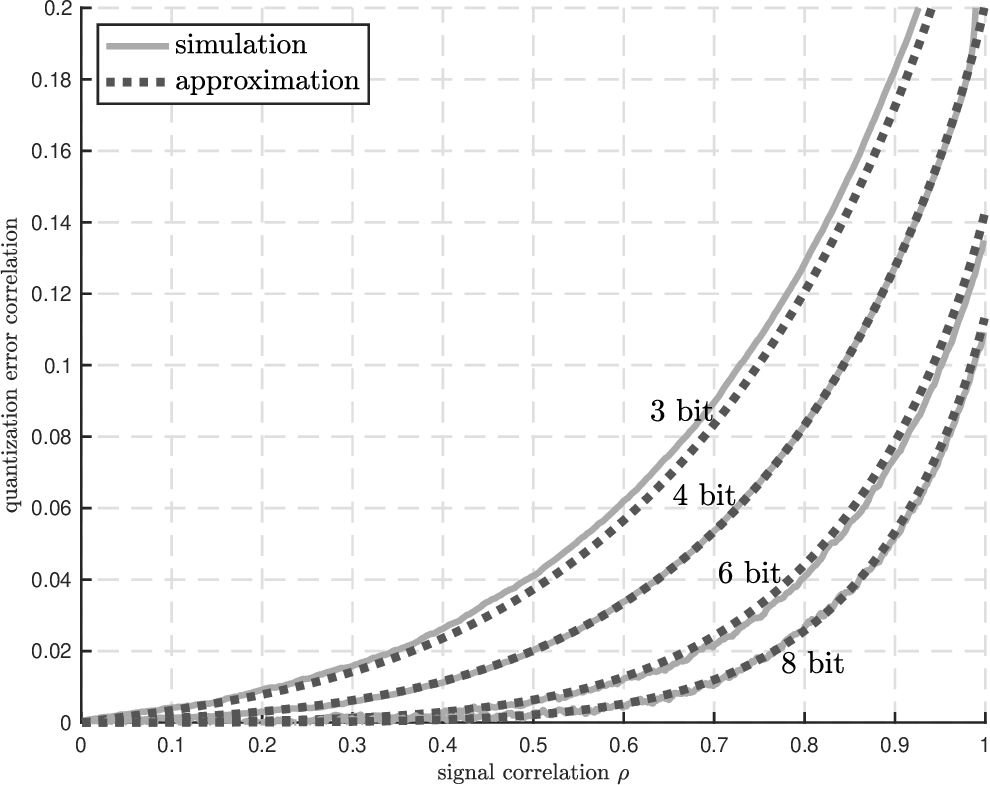}
    \caption{Simulated correlation of the quantization noise compared to the approximation (\ref{psiApprox}), normalized by $\sigma^2$ to represent the correlation.}
    \label{fig:psiSimulate}
\end{figure}
\begin{figure}
    \centering
    \includegraphics[width=.85\columnwidth]{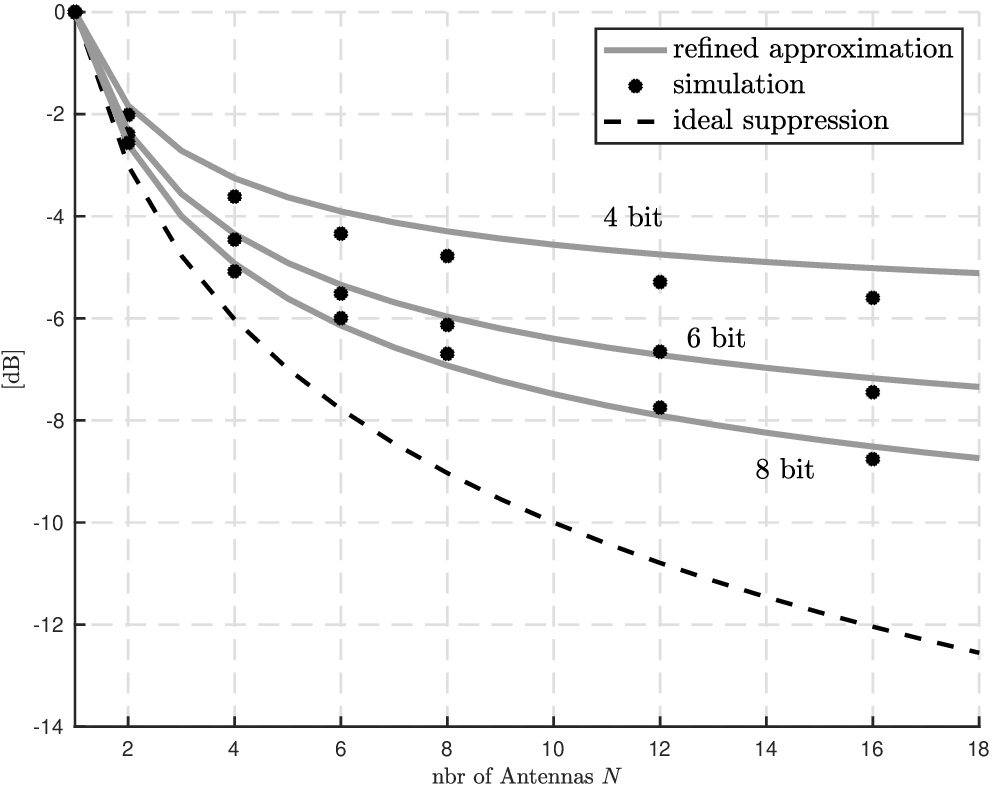}
    \caption{The noise suppression factor for various $k$. The dashed line represents the ideal noise suppression, or equivalently the minimal value of the noise suppression factor, being $\frac{1}{N}$, while the gray curves are approximations calculated by applying the more refined approximation of applying (\ref{psiApprox}) to (\ref{suppression}). The dots are noise suppression levels for simulated OFDM-transmissions.}
    \label{fig:suppSimulate}
\end{figure}
\begin{figure}
    \centering
    \includegraphics[width=.85\columnwidth]{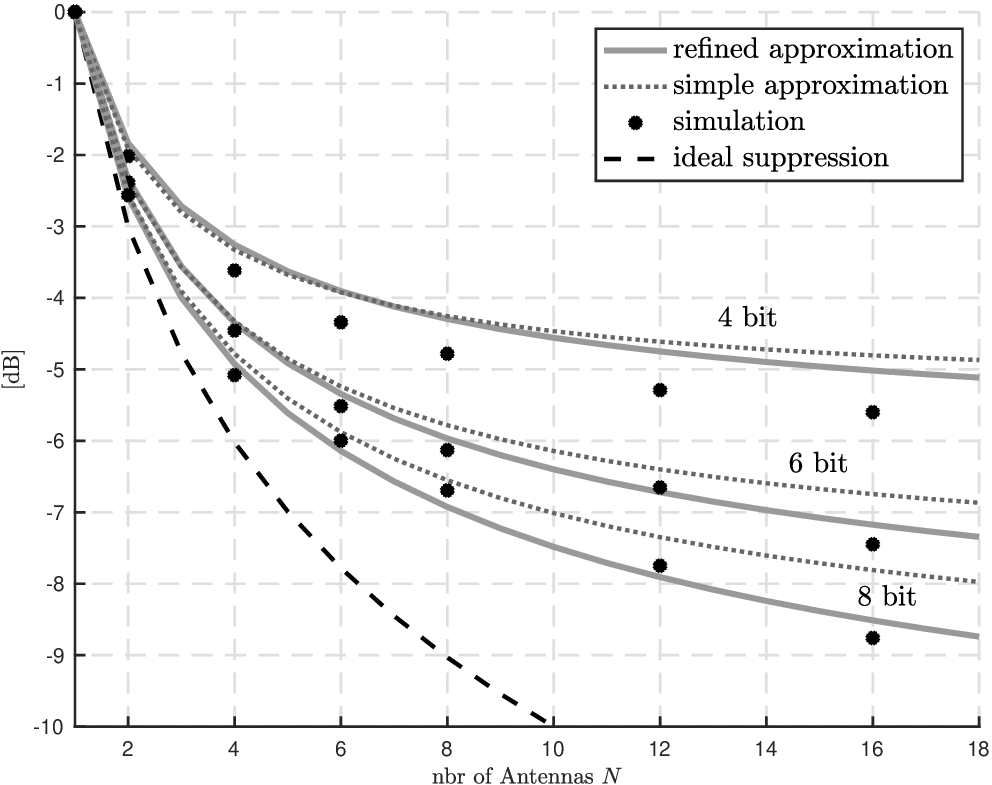}
    \caption{The comparison between the simple and refined approximation to the noise suppression.}
    \label{fig:thumbCompare}
\end{figure}
Firstly, the validity of approximating $\psi(\rho)$ as in (\ref{psiApprox}) is tested by quantizing simulated bivariate Gaussian distributed data with predefined correlation and estimating their correlation. The results in Fig. \ref{fig:psiSimulate} show that the approximation (\ref{psiApprox}) aligns well with the simulated data. Furthermore, the quantization noise correlation on the vertical axis is significantly lower than the signal correlation on the horizontal axis.

Next, QPSK-modulated OFDM-symbol transmission is simulated. The receiver has a ULA with 1-16 antennas, and we consider a uniform grid of AoA $\theta \in(-\frac{\pi}{2},\frac{\pi}{2})$ to the ULA. The ADC bit resolution is k=4-8 bits, i.e., typical numbers for a mmW mobile device implementation. In these simulations, each signal consists of $50$ OFDM-symbols with FFT-size $4096$. The ADC clipping level is set to follow the least-squares fit $ R(k) = -0.0053k^2+0.3763k+1.26 $ for the optimal clipping level, see \cite{VarianceQuant}. Numerical integration over these points yields the simulated average noise suppression factor. For each pair $k$ and $N$ for which the noise suppression is simulated, we run $4$ transmissions as described above and take the noise suppression as the sample average of these. 
The simulated average noise suppression is compared against the theoretical expression (\ref{suppression}) with $\psi(\rho)$ approximated by (\ref{psiApprox}). The suppression is shown for various numbers of bits in Fig. \ref{fig:suppSimulate}. As is readily seen therein, the accuracy of the theoretical approximation is at its worst for small $k$, since letting $\psi = \sigma^2$ for $|\rho| \leq 1-2^{-2k}$ is rather crude when $k$ is small. However, the trend is the same, so the formula serves the purpose of being a rule-of-thumb. Lastly, the accuracy of the simple approximation (\ref{upperBoundSuppression}) is tested by comparing it with the more sophisticated approach of applying (\ref{psiApprox}) to (\ref{suppression}), as shown in Fig. \ref{fig:thumbCompare}. As can be seen, for a fixed $k$-bit ADC, there is still some correlation between the respective antennas' quantization noise (the dashed  curve in the figures shows the case when the quantization noise is uncorrelated, and hence suppression is lower bounded by $ 1/N$), however, the higher bit resolution the less the correlation, and hence the larger the processing gain is achieved.

\section{Implications for Digital Phased Arrays}
\label{implications}

The quantization noise suppression is in the range of 4-9 dB when considering 5G mmW digital phased arrays relevant for mobile devices \cite{DuttaDigBFpowerCons} (i.e., 8-16 antennas and 4-8 bits per ADC). This property can be used for ADC resolution reduction.
Since each bit reduction in an ADC causes 5 dB extra quantization noise \cite{VarianceQuant}, the digital architecture allows for 1-2 bits of ADC resolution reduction compared to an ADC designed for an analog phased array architecture, where no quantization noise suppression is achieved.

The power consumption of a $k$-bit ADC is typically proportional to $2^k$ \cite{WaldenADC}.
Consider a digital phased array with 16 antennas that requires $16$ ADC pairs (I and Q), with $k$ bit resolution. This should be compared to the analog architecture's single ADC pair, where we need $k+2$ bit resolution to achieve the same quantization noise level. The ADC power consumption for the digital architecture is proportional to $ 16\cdot2^{k}$, compared to the analog architecture's $ 2^2\cdot 2^{k} $. Hence, the ADC power consumption ratio is only a factor of 4, not a factor 16 in this example.

Moreover, a digital phased array does not need analog phase shifters, which have a signal loss of 8-10 dB \cite{DuttaDigBFpowerCons} and therefore require more efficient Power Amplifiers (PA) and Low Noise Amplifiers (LNA) compensating for this loss.
Hence, the transceiver power consumption for digital phase arrays, targeting 5G mmW mobile devices may be on par or below the power consumption for corresponding analog phased arrays, as indicated in \cite{DuttaDigBFpowerCons}, while they did not take the digital phased array ADC design advantages into account in their estimates. 

\section{Conclusions}
\label{conclusions}
In this paper, we have analyzed the quantization noise characteristics for a Gaussian signal received and processed by a digital phased arrays, showing that quantization noise suppression is achieved due to the coherent combining of the signals received by the antenna array. We derived an easy to use approximate upper bound of the quantization noise suppression as a function of the number of antennas in the array and ADC bit resolution. It can be used when setting requirements on the ADC in a digital phased array transceiver design. Based on the results, we conclude that the ADC design for digital phase arrays can be relaxed compared to analog counterparts, thereby enabling the use of digital phased array architectures also in low-cost devices.

\printbibliography

\end{document}

%% file: Graphs/partitionGray.tex
\begin{figure}[t!]
    \centering
    \begin{tikzpicture}[scale = 0.4]

  \fill[pattern = north west lines, pattern color = lightgray] (-3,-3) rectangle (3,3);

  \fill[pattern = north east lines, pattern color = gray!40] (3,3) rectangle (8,8);
  \fill[pattern = north east lines, pattern color = gray!40] (-8,3) rectangle (-3,8);
  \fill[pattern = north east lines, pattern color = gray!40] (-8,-8) rectangle (-3,-3);
  \fill[pattern = north east lines, pattern color = gray!40] (3,-8) rectangle (8,-3);

  \fill[pattern = crosshatch, pattern color = gray!70] (3,-3) rectangle (8,3);
  \fill[pattern = crosshatch, pattern color = gray!70] (-3,3) rectangle (3,8);
  \fill[pattern = crosshatch, pattern color = gray!70] (-8,-3) rectangle (-3,3);
  \fill[pattern = crosshatch, pattern color = gray!70] (-3,-8) rectangle (3,-3);

  \draw[dashed] (-3,-8)--(-3,8);
  \draw[dashed] (3,-8)--(3,8);
  \draw[dashed] (-8,-3)--(8,-3);
  \draw[dashed] (-8,3)--(8,3);

  \draw[->,very thick] (-8.1, 0) -- (8.1, 0) node[right] {$X$};
  \draw[->,very thick] (0, -8.1) -- (0, 8.1) node[above] {$Y$};

  \draw[very thick] (3,-0.2) --(3,0.2) node[above] {$R$};
  \draw[very thick] (-3,-0.2) --(-3,0.2) node[above] {$-R$};
  \draw[very thick] (-0.2,3) --(0.2,3) node[above] {$R$};
  \draw[very thick] (-0.2,-3) --(0.2,-3) node[right] {$-R$};

  \node at (6,6) {$\mathcal{C}_1$};
  \node at (-6,6) {$\mathcal{C}_2$};
  \node at (-6,-6) {$\mathcal{C}_1$};
  \node at (6,-6) {$\mathcal{C}_2$};

  \node at (-1.5,2) {$\mathcal{I}$};

  \node at (-6,2) {$\mathcal{E}_3$};
  \node at (-2,6) {$\mathcal{E}_2$};
  \node at (6,2) {$\mathcal{E}_1$};
  \node at (-2,-6) {$\mathcal{E}_4$};

\end{tikzpicture}
\caption{The partition of $\mathbb{R}^2$ which the covariance is evaluated over. For legibility, $\mathcal{I}$ denotes $\bigcup_{m,n = 1}^{2^k}\mathcal{I}_{m,n}$. The edge region $\mathcal{E}$ has been partitioned further into $\mathcal{E} = \bigcup_{i =1}^{4}\mathcal{E}_i$, the reason that is explained in the appendix.}
\label{partition}
\end{figure}